\documentclass[aps,twocolumn,float,prd,psfig,amsmath]{revtex4}
\usepackage[dvipdfmx]{graphicx}
\usepackage{amsmath,amssymb,natbib,siunitx,booktabs,url}
\usepackage{color}

\newcommand{\gsim}{\gtrsim}

\newcommand{\vect}[1]{\mbox{\boldmath${#1}$}}
\newcommand{\lmk}{\left(}
\newcommand{\rmk}{\right)}

\newcommand{\lkk}{\left[}
\newcommand{\rkk}{\right]}
\newcommand{\lla}{\left\langle}
\newcommand{\p}{\partial}
\newcommand{\rra}{\right\rangle}
\newcommand{\so}{M_\odot}
\newcommand{\mch}{{\cal M}}

\newcommand{\bea}{{\begin{eqnarray}}}
\newcommand{\eea}{{\end{eqnarray}}}

\begin{document}

\title{Tracking the Long-Term GW Phase Evolution for HM Cancri-like Binaries with LISA}

\author{Naoki Seto }
\affiliation{Department of Physics, Kyoto University, 
Kyoto 606-8502, Japan
}

\date{\today}

\begin{abstract}

 From prolonged X-ray and optical data of the ultra-compact binary HM Cancri, two groups recently measured  the second derivative  of its orbital frequency.  The space gravitational wave (GW) detector  LISA will detect $\sim10^4$ Galactic binaries  and their second frequency derivatives will  be  interesting observational targets for LISA. 
Here, we forecast the GW  signal analysis  for HM Cancri, as an ideal reference system for these numerous binaries.
  We find that,  in its nominal operation period $T\sim4$yr,   LISA is unlikely to  realize a sufficient measurement precision for the reported second frequency derivative of this binary.  However, because of a strong dependence on the time baseline, the precision  will be drastically improved by extending the operation period of LISA or combining it with other missions (e.g., Taiji and TianQin) in a sequential order.

 
\end{abstract}

\maketitle

\section{introduction}

The laser interferometer spacer antenna (LISA) has potential to detect GWs from various astrophysical sources such  as merging  massive black hole binaries, extreme mass ratio inspirals and Galactic stellar mass binaries \cite{2022arXiv220306016A}. While the event rates of  the systems involving massive black holes are highly uncertain, the Galactic binaries are solid observational targets.  In fact, a few tens of  compact binaries are currently listed as LISA's verification sources and will be  detected at sufficient signal-to-noise ratios \cite{2018MNRAS.480..302K,2022arXiv220306016A}.  HM Cancri is an interacting white dwarf (WD) binary  and  regarded as one of the loudest GW sources in the list.

In this paper, since we focus on GW observation, we basically use the GW frequency $f$ of a binary, not its orbital frequency.  For a circular binary such as HM Cancri, the GW frequency  is twice of the orbital frequency. Correspondingly, if necessary, we will automatically make appropriate conversions for the reported values in the literature (often given for the orbital frequency).   
 The  GW frequency  of HM Cancri is  $f=6.2$mHz, which is the highest frequency in the list \cite{2018MNRAS.480..302K,2022arXiv220306016A}. 

Using electromagnetic (EM) data accumulated for HM Cancri in the past $\sim20$ years, two groups recently  measured the acceleration rate of its frequency $\ddot f$, for the first time among the verification binaries \cite{2021ApJ...912L...8S,2023MNRAS.518.5123M}.  The second derivative  $\ddot f$ is expected  to contain interesting information on the evolution of the binary  (e.g., mass transfer \cite{2006ApJ...649L..99D,2006ApJ...653.1429D,2012ApJ...758...64K,2021ApJ...912L...8S,2023MNRAS.518.5123M}, see also \cite{Biscoveanu:2022sul} for $\dot f$) and its environment (e.g., tertiary gravitational perturbation \cite{2018PhRvD..98f4012R,seto233}).

  Owing to its omni-directional sensitivity, LISA will detect $\sim 10^4$ Galactic binaries, including more than $\sim 10^2$ subset at $f\gsim  6$mHz \cite{2004MNRAS.349..181N,2022arXiv220306016A,Lamberts:2019nyk,2022PhRvL.128d1101S}. 
 The obtained results $\ddot f$ for HM Cancri will serve as ideal reference values, when discussing GW signal analysis for these numerous binaries. 
  In this paper, we study such an observational prospect, paying a special attention to HM Cancri. 
  Relatedly, we assort useful analytical expressions  based on the Fisher matrix approach.

  As one can easily expect, the measurement  error for  $\ddot f$ depends strongly on the time baseline of the observation. Therefore, our study will be useful for designing the mission lifetime of LISA  or coordinating  its collaboration with other detectors such as Taiji \cite{2018arXiv180709495R} and TianQin \cite{2016CQGra..33c5010L}.

This paper is organized as follows.  In Sec. II, we briefly summarize  the recent long-term orbital analysis for HM Cancri.  We also mention some astrophysical  implications of the observed results. In Sec. III, applying the Fisher matrix  approach to a simplified phase model, we analytically evaluate the estimation  errors for phase related parameters such as $\dot f$ and $\ddot f$.   In Sec. IV, we examine the validity of our analytical expressions, by comparison  with the full Fisher matrix predictions including all the  fitting   parameters.    In Sec. V, we discuss the observational prospects of measuring $\ddot f$ for HM Cancri with LISA. We also apply our analytical expressions to two other verification binaries. In Sec. IV,  we discuss issues related to our study.  Sec. VII is a short summary of this paper.

\section{long-term orbital evolution of HM Cancri}

\subsection{Observed Results}

Using X-ray data accumulated  in the past $\sim 20$ years, Strohmayer \cite{2021ApJ...912L...8S} examined the long-term orbital evolution  of HM Cancri. He fitted  its phase evolution with the following cubic function 
\begin{equation}
\Phi(t)=2\pi \lmk ft+\frac{{\dot f}t^2}{2!}+\frac{{\ddot f}t^3}{3!}   \rmk+\varphi  \label{cub}
\end{equation}
with the time origin $t=0$ at a certain epoch in January 2004.
The fitted parameters  (after the aforementioned conversions) are  
\begin{eqnarray}
f&=&0.0062 {\rm Hz},\\
{\dot f}&=&(7.11\pm  0.01)\times 10^{-16} {\rm Hz\, s^{-1}},\label{ch1}\\
{\ddot f}&=&(-1.79\pm0.28)\times 10^{-26}{\rm Hz\,s^{-2}}. \label{dds}
\end{eqnarray}
Here the error bars represent the $1\sigma$ uncertainties.

Munday et al. \cite{2023MNRAS.518.5123M} analyzed HM Cancri's optical data with a baseline  of  $\sim20$ years and found  a good fit to the cubic functional form (\ref{cub}). Their fitted second derivative $\ddot f$ is 
 \begin{eqnarray}
{\ddot f}=(-1.08\pm0.42)\times 10^{-26}{\rm Hz\,s^{-2}}. \label{dds2}
\end{eqnarray}
Note that the error bars in Eqs. (\ref{dds}) and (\ref{dds2})  are nearly overlapped.  Below, we mainly use Eq. (\ref{dds}) as a reference value of $\ddot f$.

\subsection{Astrophysical Implications}

Before studying the parameter estimation errors  with LISA, we briefly discuss some astrophysical implications of the observed long-term orbital  evolution  of HM Cancri.

As mentioned in the previous subsection, it took $\sim20$ years to resolve  the second frequency derivative $\ddot f$ for HM Cancri.  In contrast,  shortly after the identification of this binary,  its chirp rate $\dot f$  was measured at a value close    to Eq. (\ref{ch1}) \cite{2004MSAIS...5..148I,2005ApJ...627..920S}.  If the gravitational radiation reaction  dominates  the observed chirp rate $\dot f$, we should have 
 \begin{eqnarray}
 {\dot f}\simeq{\dot f}_{\rm R}&=&\frac{96\pi^{8/3}G^{5/3} f^{11/3}\mch^{5/3}}{5c^5}\\
 &=&  7.0 \times 10^{-16} \lmk  \frac{f}{\rm 6.2mHz}  \rmk^{11/3} \lmk  \frac{\mch}{0.32\so}  \rmk^{5/3}.
 \end{eqnarray}
  Given this relation and the observed rate $\dot f$,  the chirp mass of HM Cancri was roughly estimated  to be {$\mch\sim 0.3\so$} \cite{2021ApJ...912L...8S,2023MNRAS.518.5123M}. 
If this binary is  mainly evolved  by the gravitational radiation reaction, we also have ${\ddot f}\simeq {\ddot f}_R= 11{\dot f}^2/(3f)\sim 1.5 \times 10^{-28}{\rm Hz\,s^{-2}}$ as predicted long before the actual measurement  of $\ddot f$ \cite{2006ApJ...649L..99D}. Interestingly, the observed result (\ref{dds})  is totally different from the predicted one ${\ddot f}_{\rm R}$.

Using the {Modules for Experiments in Stellar Astrophysics} (MESA) code \cite{2019ApJS..243...10P}, 
 Munday et al. \cite{2023MNRAS.518.5123M} compared the observed values $({\dot f},{\ddot f})$ with the simulated frequency evolution of many  WD binaries.    They pointed out that HM Cancri might be shortly ($\sim 10^3$yr) before the frequency maximum and discovered with the help of  a selection effect (see also \cite{2021ApJ...912L...8S}).

On another front, observations suggest that a significant fraction of  close white dwarf binaries might be in triple or higher-order  multiple systems \cite{2017A&A...602A..16T} (see also \cite{2018PhRvD..98f4012R,2021MNRAS.502.4199X} in the context of LISA observation). In \cite{seto233}  the author argued the possibility  that HM Cancri has  a tertiary component.   For  a circular outer orbit with a period $P_3$ (much longer than the observation period),  the tertiary perturbation  generates  the following shifts to the frequency derivatives (see  \cite{2016ApJ...826...86K,2016MNRAS.460.2207B} for similar effects on pulsar timing analysis) 
\begin{eqnarray}
 {\dot f}_{3}&=&1.0\times 10^{-16} F\cos\varphi_3 \lmk \frac{f}{\rm 6.2mHz} \rmk  \lmk \frac{M_T}{2\so} \rmk^{1/3}\nonumber\\
 & &\times  \lmk \frac{P_3}{\rm 250yr} \rmk^{-4/3} \,{\rm Hz\,s^{-1}}\label{t1}\\
  {\ddot f}_{3}&=&-8.0\times 10^{-26} F  \sin\varphi_3\lmk \frac{f}{\rm 6.2mHz} \rmk  \lmk \frac{M_T}{2\so} \rmk^{1/3}\nonumber\\
 & &\times  \lmk \frac{P_3}{\rm 250yr} \rmk^{-7/3}\,{\rm Hz\,s^{-2}}.\label{t2}
 \end{eqnarray}
Here $M_T$ is the total mass of the triple system, $\varphi_3$ is the outer orbital phase, and $F=m_3 \sin I_3/M_T$  is a projection factor with the tertiary mass  $m_3$ and the outer inclination angle  $I_3$.   Therefore a dark tertiary  component (e.g., an old WD with an outer orbital period of $\sim250$ years) can  serves as the main cause for the observed value  $\ddot f$, with a limited impact on the observed first derivative $\dot f$ (see  Eqs. (\ref{ch1})(\ref{dds})(\ref{t1}) and (\ref{t2})) \cite{seto233}.

As mentioned earlier, during its operation period, LISA is expected to detect $\sim 10^4$ Galactic compact binaries \cite{2022arXiv220306016A}. In particular, it will make a complete Galactic  survey for binaries at $f\gsim 5$mHz \cite{2022arXiv220306016A}.  
 Many of the detected binaries  might have intensified rates $\ddot f$ due to their tertiaries. If this  is the case,    the probability distribution  function of the observed values $\ddot f$ would be  nearly symmetric  around the origin ${\ddot f}=0$.

\section{Simplified  phase model}

In this section, applying the Fisher matrix approach to a simplified GW phase model,  we analytically estimate the measurement errors for the phase related parameters such as $\dot f$ and $\ddot f$ (see also \cite{2002MNRAS.333..469S,2018PhRvD..98f4012R,2014ApJ...791...76S}).   

\subsection{Basic Prescription}

Here the GW phase is assumed to be  well described  by  the following Taylor expansion with the coefficients  $\{f,{\dot f},{\ddot f}\}$ defined at the time origin $t=0$
\begin{eqnarray}
h(t)&=& A\cos[\Phi(t)]\\
&=&A\cos\lkk 2\pi \lmk ft+\frac{{\dot f}t^2}{2!}+\frac{{\ddot f}t^3}{3!}   \rmk+\varphi \rkk. \label{wf}
\end{eqnarray}

We consider a  signal analysis in the time interval $t\in[t_1,t_1+T]$ with the initial time $t_1$ and the observational duration $T$.  From  Parceval's theorem for a  nearly monochromatic waveform \citep{1998PhRvD..57.7089C}, the signal-to-noise ratio $\rho$ is  evaluated as 
\begin{eqnarray}
\rho^2&=&\frac2{S_{\rm n}(f)}\int_{t_1}^{t_1+T}  h(t)h(t)dt \label{inner}\\
&=&\frac{A^2 T}{S_{\rm n}(f)}  \label{snr}
\end{eqnarray}
with the measurement  noise spectrum   $S_{\rm n}(f)$. 
Note that, as long as the GW signal is nearly monochromatic  (i.e. ${|\dot f}| T\ll f$ and  ${|\ddot f}| T^2\ll f$), 
we can effectively set the true values at  ${\dot f}=0$ and  ${\ddot f}=0$ for   evaluating the inner product (\ref{inner}) (also  for the Fisher matrix elements below).

Next we apply the Fisher matrix approach  to the phase related parameters  ${\vect \theta}=\{\varphi, f,{\dot f},{\ddot f}\}$ with  the waveform model (\ref{wf}).  The amplitude parameter $A$ has no correlation with the target parameters  ${\vect \theta}$ and can be dropped from our fitting parameters in this section.

We can formally express the Fisher matrix elements as 
\begin{eqnarray}
F_{\theta_i\theta_j}=\frac{2}{S_{\rm n}(f)}\int_{t_1}^{t_1+T} \p_{\theta_i }h(t) \p_{\theta_j}h(t)  dt.
\end{eqnarray}
Using Eq. (\ref{snr}),  we can eliminate the overall  factor $A/S_{\rm n}(f)$ and then present the matrix elements $F_{\theta_i\theta_j}$ in terms of  $\rho$, $t_1$ and $T$. For example, we have
\begin{eqnarray}
F_{ff}=4\pi^2 \rho^2  \lmk  \frac{3t_1^2+3t_1 T+T^2}3 \rmk.
\end{eqnarray}

After taking the inverse of the matrix  $F$, we can evaluate  the covariance matrix of the parameter estimation  error  $ \delta{\theta_i}$ as 
\begin{eqnarray}
\lla \delta{\theta_i} \delta{\theta_j}  \rra=(F^{-1})_{\theta_i\theta_j}.
\end{eqnarray}

Below, for notational simplicity, we put the rms errors by 
\begin{eqnarray}
\Delta{\theta_i}\equiv \lla \delta{\theta_i} \delta{\theta_i}  \rra^{1/2},
\end{eqnarray}
and denote the correlation factor by
\begin{eqnarray}
 C_{\theta_i\theta_j}\equiv \frac{\lla \delta{\theta_i} \delta{\theta_j}  \rra}{\lla \delta{\theta_i} \delta{\theta_i}  \rra^{1/2} \lla \delta{\theta_j}\delta{\theta_j} \rra^{1/2}}
\end{eqnarray}
with  the Cauchy-Schwartz inequality
$|C_{\theta_i\theta_j}|\le 1$.  

\subsection{Expansion at the Initial  Epoch}
Following the outline in the previous subsection, we now evaluate the Fisher matrix $F$   for the observational time domain $t\in [0,T]$ with the initial epoch at $t_1=0$.  This means that  the GW phase is Taylor expanded  as in Eq. (\ref{wf}) at the initial epoch $t=t_1=0$.   For the four basic parameters ${\vect \theta}=\{\varphi, f,{\dot f},{\ddot f}\}$,  we can easily obtain the rms errors as
\begin{eqnarray}
\Delta\varphi=\frac4\rho,~\Delta f=\frac{10\sqrt3}{\pi \rho T}, ~\Delta {\dot f}=\frac{36\sqrt5}{\pi \rho T^2},~\Delta {\ddot   f}=\frac{60\sqrt7}{\pi \rho T^3} . \label{v1}
\end{eqnarray}
We provide some examples of the correlation  coefficients
\begin{eqnarray}
C_{{\dot f}{\ddot f}}=-\frac{\sqrt{35}}6\simeq -0.986,~~C_{{ f}{\ddot f}}=-\frac{\sqrt{21}}5\simeq -0.917.
\end{eqnarray}
Robson et al. \cite{2018PhRvD..98f4012R} obtained essentially the same results as Eq. (\ref{v1}), while  their definitions for the evolutionary parameters  $({\dot f},{\ddot f})$ are slightly different from ours (see their appendix). 

Next, for a comparison purpose,  we drop the cubic term $\propto {\ddot f}t^3$ in Eq. (\ref{wf}) as
\begin{equation}
h_{\rm tr}(t)=A\cos\lkk 2\pi \lmk ft+\frac{{\dot f}t^2}{2!}   \rmk+\varphi \rkk. \label{wf2}
\end{equation}
Using this truncated waveform model studied in \cite{2002MNRAS.333..469S}, 
 we  can derive the analytical results for the smaller set  ${\vect \theta}=\{\varphi, f,{\dot f}\}$ as follows
\begin{eqnarray}
\Delta\varphi=\frac3\rho,~\Delta f=\frac{4\sqrt3}{\pi \rho T}, ~\Delta {\dot f}=\frac{6\sqrt5}{\pi \rho T^2}, \label{v11}
\end{eqnarray}
which are identical to the corresponding expressions in \cite{2002MNRAS.333..469S}.
Compared with Eq. (\ref{v1}), we can see a large reduction of the error $\Delta {\dot f}$. 

\subsection{Expansion at the Mid Point}
Next we evaluate the Fisher matrix prediction  for the observational time domain $t\in[-T/2,T/2]$, setting the initial epoch at
$t_1=-T/2$.  In  this choice, the GW phase is Taylor expanded  as Eq.  (\ref{wf})  at the midpoint ($t=0$) of the observational period. The estimation errors for the parameters $\{\varphi_0, {\dot f}\}$ and those for   $\{f, {\ddot f}\}$ are statistically uncorrelated, and the matrices $F$ and $F^{-1}$ are block diagonal.  This is because we  have the  symmetric cancellations at integrating odd functions such as 
\begin{eqnarray}
F_{{\ddot f}{\dot f}}\propto \int_{T/2}^{-T/2} t^{2+3}dt=0.
\end{eqnarray}

For the fitting parameters ${\vect \theta}=\{\varphi, f,{\dot f},{\ddot f}\}$, the rms errors are estimated to be 
\begin{eqnarray}
\Delta\varphi=\frac3{2\rho},~\Delta f=\frac{5\sqrt3}{2\pi \rho T}, ~\Delta {\dot f}=\frac{6\sqrt5}{\pi \rho T^2},~\Delta {\ddot   f}=\frac{60\sqrt7}{\pi \rho T^3} \label{v2}
\end{eqnarray}
with some examples for the correlation coefficients as 
\begin{eqnarray}
C_{{\dot f}{\ddot f}}=0,~~C_{{ f}{\ddot f}}=-\frac{\sqrt{21}}5.
\end{eqnarray}
As in the case of Eq.  (\ref{v11}),   the magnitude  $\Delta {\dot f}$ is largely reduced from Eq. (\ref{v1}).   In contrast,  the results   for  $\Delta {\ddot f}$ are the same in Eqs. (\ref{v1}) and (\ref{v2}). We can understand this from  the following argument.   Given the difference between the time origins ($t_1=0$ or $-T/2$), we can relate the parameters $\{\varphi, f,{\dot f},{\ddot f}\}$ in this subsection with those  $\{\varphi_{\rm o}, f_{\rm o},{\dot f}_{\rm o},{\ddot f}_{\rm o}\}$ (temporarily attaching the subscript ``$\rm o$'') in the previous subsection. For example, we  have 
\begin{eqnarray}
{\dot f}={\dot f}_{\rm o}+{\ddot f}_{\rm o}T/2, ~~{\ddot f}={\ddot f}_{\rm o},  \label{rele}
\end{eqnarray}
resulting in  $\Delta {\ddot f}=\Delta {\ddot f}_{\rm o}$  for  optimal signal  analyses.  

We again examine the truncated model  (\ref{wf2}) 
and obtain the associated measurement  errors under the mid point expansion as
\begin{eqnarray}
\Delta\phi=\frac3{2\rho},~\Delta f=\frac{\sqrt3}{\pi \rho T}, ~\Delta {\dot f}=\frac{6\sqrt5}{\pi \rho T^2}.\label{v22}
\end{eqnarray}
Because of  the block diagonalization, the magnitudes $\Delta \phi$ and $\Delta {\dot f}$  are the same as Eq. (\ref{v2}).  Furthermore, due to the reason similar to Eq. (\ref{rele}), we have the same expression $\Delta {\dot f}$ in Eqs. (\ref{v11})  and (\ref{v22}).

\section{Comparison with Full Fisher Matrix Analysis}

Our simplified  analytical models in the previous section  lack the amplitude and  Doppler phase modulations, which are induced by the annual motion of   LISA \citep{1998PhRvD..57.7089C}.  
In this subsection, we examine the validity of our simple analytical results (\ref{v1}) and (\ref{v2}), by comparing them  with the full Fisher matrix results evaluated numerically.

For the full Fisher  matrix estimation, we deal with the following nine fitting parameters $\{A, \varphi, f,{\dot f},{\ddot f},\theta_S,\phi_S,\theta_L,\phi_L\}$,  extending  the formulation in  \citep{1998PhRvD..57.7089C}.  The  pairwise angular parameters $\{\theta_S,\phi_S\}$ and $\{\theta_L,\phi_L\}$ respectively specify the source direction and orientation in the ecliptic coordinate.

Here we comment on an earlier study related to this section. 
Takahashi and Seto \cite{2002ApJ...575.1030T} briefly   compared  the analytical results (\ref{v11}) (for the truncated model (\ref{wf2})) with the corresponding numerical results.  The latter were obtained  from the full $8\times8$ Fisher matrices  (without the parameter $\ddot f$).  They  found that, for $T\gtrsim  2$yr, the analytical and numerical results agree quite well. This is because, for  $T\gtrsim  2$yr,  the intrinsic frequency evolution in Eq. (\ref{wf2}) is well separable from the annual periodic phase shift induced by LISA's motion. 
Even adding the cubic term as  Eq. (\ref{wf}), we can expect a similar trend.

For numerically evaluating the $9\times9$ Fisher matrix, 
we randomly generated  20 realizations for the combinations  $\{\theta_S,\phi_S,\theta_L,\phi_L\}$, assuming that the direction and orientation vectors are isotropically distributed. 
At $f=1$mHz and 10mHz,  we numerically evaluated the estimation errors $ (\Delta{\dot f}_{\rm full},\Delta{\ddot f}_{\rm full})$ and  compared them with the analytical ones $(\Delta{\dot f}_{\rm ana},\Delta{\ddot f}_{\rm ana})$ given  in the previous section.   Only in this section, we use the subscripts ``full'' and ``ana' respectively for  the full Fisher matrix estimations and the analytical ones.   

In Fig. 1, for $T=1,2,4,6$ and 8yr, we present the ratio $ {\Delta{\ddot f}_{\rm full}}/{\Delta{\ddot f}_{\rm ana}}$ for the second derivative $\ddot f$ with a small horizontal offset for the mid point expansion. As shown in 
Eqs. (\ref{v1}) and (\ref{v2}), the analytical expression ${\Delta{\ddot f}_{\rm ana}}={60\sqrt7}/{(\pi \rho T^3)}$ is identical in the two expansion methods. Consistent with the argument around Eq. (\ref{rele}),  Fig. 1 clearly shows that the numerical results ${\Delta{\ddot f}_{\rm full}}$ are also  the same for  the two  expansion methods (except for tiny numerical errors).

At $T\gsim  2$yr, the mismatches between the two estimations ${\Delta{\ddot f}_{\rm full}}$ and ${\Delta{\ddot f}_{\rm ana}}$ are less than 50\%.  In reality,  a long-term signal integration is essential for measuring the second derivative $\ddot f$, and the mismatches are even less than $\sim 5$\% at $T\gsim 8$yr. 

In contrast, for $T=1$yr, the analytical estimation ${\Delta{\ddot f}_{\rm ana}}$ underestimates the full estimation ${\Delta{\ddot f}_{\rm full}}$, generally showing  larger mismatches at $f=$10mHz.  This frequency dependence can be understood from the relative importance of the amplitude modulation for estimating the source direction. With a LISA-like detector, we can determine the source direction from the Doppler phase modulation and the amplitude modulation, both induced by the annual motion of the detector  \citep{1998PhRvD..57.7089C}.  If  the signal-to-noise ratio is fixed, the former is proportional to the frequency but the latter is independent of  it. 
 Therefore, at the lower frequency regime, the amplitude modulation can work more efficiently to suppress the interference between  the Doppler phase modulation and the intrinsic phase evolution. 

In their appendix, Robson et al.  \cite{2018PhRvD..98f4012R} derived an analytical expression (essentially corresponding to ${\Delta{\ddot f}_{\rm ana}}$ in Eq. (\ref{v1})).  They also  compared it with a full Fisher matrix prediction.  For a certain set of the geometric parameters  $\{\theta_S,\phi_S,\theta_L,\phi_L\}$ at $f=5$mHz, they reported a mismatch of $ {\Delta{\ddot f}_{\rm full}}/{\Delta{\ddot f}_{\rm ana}}\sim  6$.  While they did not explicitly   present the applied integration period $T$,  our Fig.  1 indicates that their choice is  likely to be $T\sim1$yr. 

In Fig. 2, we show the ratio $ {\Delta{\dot f}_{\rm full}}/{\Delta{\dot f}_{\rm ana}}$ for the first derivative $\dot f$.  Now, the analytical results $\Delta {\dot f}_{\rm ana}$ are different between the two expansion methods (compare Eqs. (\ref{v1}) and (\ref{v2})).  In any case, we can again confirm that, at $T\gsim2$yr,  the analytical results $ {\Delta{\dot f}_{\rm ana}}$ well reproduce the numerical ones  $ {\Delta{\dot f}_{\rm full}}$.

The ratios in  Figs. 1 and 2 are  slightly less  than  unity for some binary samples, even  though the number of fitting parameters are larger for the numerator. This is actually not surprising, given the time dependence of the signal accumulation. In contrast to the flat weighting  for the analytic model as in Eq. (\ref{inner}), the full Fisher matrix evaluation has the annual amplitude modulation. Considering the advantage  of a longer  baseline at estimating the variation rates $\dot f$ and $\ddot f$,  the errors $({\Delta{\dot f}_{\rm full}},{\Delta{\ddot f}_{\rm full}})$ can be smaller for  a binary whose signal strength  is  relatively large around the initial and the final epochs.

At $T\gtrsim2$yr,  the simple analytic expressions $({\Delta{\dot f}_{\rm ana}},{\Delta{\ddot f}_{\rm ana}})$ work well,  even if we remove the source direction angles $\{\theta_S,\phi_S\}$ from our fitting parameters (e.g., after identifying the EM counterpart).  We can easily understand  this from the weak   correlation between the intrinsic phase parameters and the direction angles  at $T\gtrsim2$yr.  

 We  should also note  that  the full Fisher matrix predictions  still have  some deviations from more elaborate evaluations such as Markov Chain Monte Carlo simulations \cite{2018PhRvD..98f4012R} (see also \cite{2008PhRvD..77d2001V}).  Nevertheless,  the Fisher matrix predictions  (in particular, our analytical expressions) will be  convenient guides for discussing parameter estimation  errors.

\begin{figure}
 \includegraphics[width=.95\linewidth]{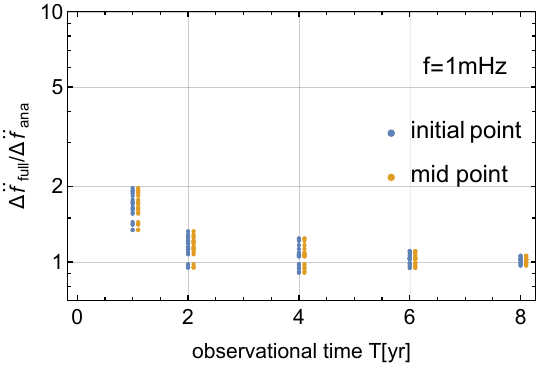} 
  \includegraphics[width=.95\linewidth]{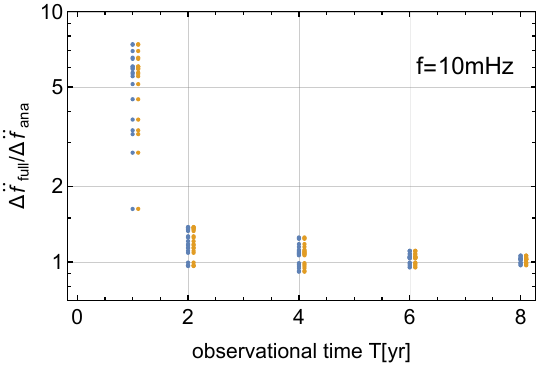}
 \caption{Ratios  ${\Delta{\ddot f}_{\rm full}}/{\Delta{\ddot f}_{\rm ana}}$ between the  full Fisher matrix predictions ${\Delta{\ddot f}_{\rm full}}$ and those with the  analytical estimation ${\Delta{\ddot f}_{\rm ana}}={60\sqrt7}/{(\pi \rho T^3)}$ given in Eqs. (\ref{v1}) and (\ref{v2}).  We show the results for  20 randomly sampled binaries with the initial and mid point expansions.  The upper panel is  for $f=1$mHz and the lower one for 10mHz.   
 }  \label{fig:volume}
\end{figure}

\begin{figure}
 \includegraphics[width=.95\linewidth]{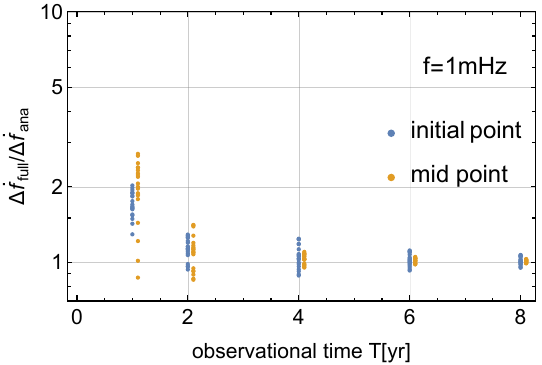} 
  \includegraphics[width=.95\linewidth]{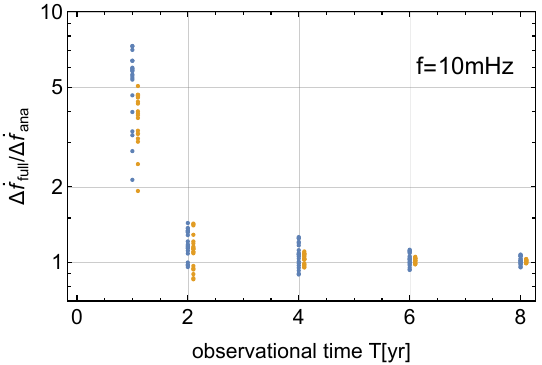}
 \caption{Similar to Fig. 1 but  for the ratio  ${\Delta{\dot f}_{\rm full}}/{\Delta{\dot f}_{\rm ana}}$.  Note, that,  between the two expansion methods,   the denominator ${\Delta{\dot f}_{\rm ana}}$ is different  by a factor of six,   as presented  in Eqs. (\ref{v1}) and (\ref{v2}).
 }  \label{fig:volume}
\end{figure}

\section{Observational prospects  with LISA}
Now, for  existing binaries such as HM Cancri, we discuss the prospects of measuring $\dot f$ and  $\ddot f$ with LISA.  We apply our analytical expressions derived in Sec. III.
\subsection{HM  Cancri}

The distance $d$ to HM Cancri has large uncertainties \citep{2021ApJ...912L...8S,2023MNRAS.518.5123M}.
Below, we use the reference  value $d=5$kpc, following \cite{2018MNRAS.480..302K}.  Then, including geometrical parameters,  LISA will detect its quadrupole GW at the signal-to-noise ratio of 
\begin{eqnarray}
\rho=210 \lmk  \frac{T}{\rm 4yr}\rmk^{1/2}   \lmk \frac{d}{\rm 5kpc} \rmk^{-1} \lmk \frac{\mch}{ 0.33\so} \rmk^{5/3}.\label{snrh}
\end{eqnarray}

Now we discuss the parameter estimation errors for the frequency derivatives $\dot f$  and $\ddot f$.  Here we solely use the expressions in Eq. (\ref{v2})  for the mid point expansion. 
Note that, as mentioned earlier,  for  the first derivative, the expression  $\Delta{\dot f}=6\sqrt{5}/(\pi \rho T^2)$ in Eq. (\ref{v2}) is identical to those in  Eqs. (\ref{v11}) and (\ref{v22}),  which are given for the truncated model   (\ref{wf2}). Therefore, we can apply this expression under various situations.

From Eqs. (\ref{v2})   and (\ref{snrh}),  at $T\gsim2$yr, we can expect the estimation errors for HM Cancri as follows
\begin{eqnarray}
\Delta {\dot f}=1.3 \times 10^{-18}  \lmk\frac{\rho}{210}   \rmk^{-1} \lmk  \frac{T}{\rm 4yr}\rmk^{-2}{\rm Hz\,s^{-1}}, \label{s0}\\
\Delta {\ddot f}=1.2 \times 10^{-25}  \lmk\frac{\rho}{210}   \rmk^{-1} \lmk  \frac{T}{\rm 4yr}\rmk^{-3}{\rm Hz\,s^{-2}}. \label{st}
\end{eqnarray}

Now let us  compare these expressions with the actual observational results (\ref{ch1}) and  (\ref{dds}).
The nominal operation period  of LISA is planned to be  $T\sim$4yr with a possible extension to $\sim$10yr.  For $T\sim 4$yr,  we can realize a good resolution $\Delta {\dot f}$ comparable to   the error bar in Eq.  (\ref{ch1}).

Meanwhile,  for $T\sim 4$yr, we can only set a loose bound to the observed value  $\ddot f$ (i.e. $\Delta {\ddot f}>|{\ddot f}|$). 
However,    given the strong scaling relation 
$\Delta {\ddot f} \propto \rho^{-1}T^{-3}\propto  T^{-7/2}$ (valid for $T\gsim 2$yr), we should not be too pessimistic.   Indeed, compared with $T=4$yr, the resolution  $\Delta {\ddot f}$ will be improved by a factor of   $\sim25$ and $\sim 100$  respectively for $T=10$ and  15yr.  Therefore, with these extended time baseline, we can reach 
$\Delta {\ddot f} =0.72\times 10^{-26}{\rm Hz\,s^{-2}}$  and  $0.18\times 10^{-26}{\rm Hz\,s^{-2}}$, corresponding to 40\% and 10\% of the reference value (\ref{dds}). 

{Note that LISA also allows us to estimate the intrinsic GW amplitude ${\cal A}\simeq \mch^{5/3}/d$ with  the typical accuracy of $\Delta {\cal A}/{\cal A}\sim 0.2 (\rho/10)^{-1}$ \cite{2002ApJ...575.1030T}.
If the observed chirp rate $\dot f$ is dominated by the radiation reaction as in Eq. (6), we can estimate the chirp mass $\mch$ and thereby  the source distance $d$.  However,  HM Cancri is an interacting binary, and we should be careful to apply Eq. (6) to this system. In  any case, using certain priors to its chrip mass distribution, we will be able to constrain its distance $d$ much better than the current estimation.  }

\subsection{V407 Vul and SDSS J0651}

Here we apply Eqs. (\ref{s0}) and (\ref{st})  to two other well-known verification binaries (see \cite{2018MNRAS.480..302K} for their basic parameters).

V407 Vul is likely to   be an interacting WD binary at the estimated distance of $d=1.8$kpc.  Its GW frequency is  $f=3.5$mHz with the observed chirp rate ${\dot f}=2.0\times 10^{-17}{\rm Hz\,s^{-1}}$. After a 4yr integration, LISA is expected to observe its GW at $\rho=170$.   From Eq. (\ref{s0}), at $T=4$yr,  we will have the resolution ${\dot f} /\Delta {\dot f}=12$.   If we can operate LISA for 10 years, the tertiary perturbation  ${\ddot f}_3$ in Eq. (\ref{t2}) can be resolved at $3\sigma$-level (i.e. $|{\ddot f}_3|/\Delta {\ddot f}>3$)  for an outer orbital period of $P_3<280$yr.  Here we put $F=0.5$,  $M_T=2.0\so$ and $\sin  \varphi_3=1$  in Eq. (\ref{t2}).  For $T=15$yr, we obtain $P_3<500$yr.

SDSS\,J0651 is a detached WD binary at $d=0.93$kpc with $f=2.6$mHz and ${\dot f}=3.3\times 10^{-17}{\rm Hz\,s^{-1}}$. For $T=4$yr, the expected signal-to-noise ratio is $\rho=90$ with the resolution  ${\dot f} /\Delta {\dot f}=11$.  For $T=10$yr, a tertiary perturbation ${\ddot f}_3$ can be resolved at $3\sigma$-level for  $P_3<180$yr.  {Since this system is a detached binary,  its distance $d$ will be estimated relatively  well, only from GW  observation. }

\section{Discussions}

\subsection{Other Detectors}

So far, we have focused on observation with LISA.  Around  4-20mHz, the Chinese proposal Taiji is planned to have $\sim 1.5$ better   sensitivity $\sqrt{S_{\rm n}(f)}$ than LISA \citep{2018arXiv180709495R}. Moreover,   HM Cancri is a very special target for another Chinese project TianQin \citep{2016CQGra..33c5010L}. Indeed, the orbital configuration of  TianQin is  designed  to optimally detect this binary \cite{2016CQGra..33c5010L}.
By  combining these missions with LISA, we might effectively realize a long time baseline of $T\gsim 15$yr and finely measure $\ddot f$ for many binaries. 
 The Japaneses projects  B-DECIGO and DECIGO are planned to explore the 0.1Hz band \citep{2001PhRvL..87v1103S,2021PTEP.2021eA105K} and can also contribute to this observational campaign.

\subsection{Relation  to EM Observations}

In the previous  section, we made a case study for HM  Cancri, as a representative target for LISA.  Now let us suppose that LISA actually  observes HM Cancri from 2035 to 2050 ($T\sim15$yr).  At the time of 2050, its EM data have the total duration of $\sim 50$yr (from $\sim 2000$)  and the resultant resolution $\Delta {\ddot f}$ would be much better than that from the GW observation. Nevertheless, the direct comparison between the EM and GW results will be  meaningful for HM Cancri. In  particular, GW emission  in itself is robustly related to the binary orbital motion with basically no need for internal dissipation models.

\section{summary}

Recently, for HM Cancri,  Strohmayer \cite{2021ApJ...912L...8S} and Munday et al. \cite{2023MNRAS.518.5123M} measured the second frequency derivative  ${\ddot f}\sim -10^{-26}{\rm Hz\,s^{-2}}$, using X-ray and optical data accumulated in the past  $\sim 20$ years. Their results will be  good  reference values  for numerous Galactic binaries to be detected by LISA. 

In this paper, based on a simplified phase model, we present analytical expressions for the estimation errors of the phase related parameters.  Our analytical expressions work well  for an observational  period $T\gsim2$yr. For HM Cancri,  LISA is unlikely to realize a sufficient resolution $\Delta {\ddot f}$ in its nominal operation period 4yr. 
However, because of the strong scaling relation $\Delta {\ddot f}\propto T^{-7/2}$, we can make much better resolution, by extending the operation period of  LISA or combining it with other detectors in a sequential order.

\section*{Acknowledgements}
The  author would like to thank N. Cornish  for correspondence. 
This work is supported by JSPS Kakenhi Grant-in-Aid for  Scientific Research (Nos.~   19K03870 and 23K03385).

\bibliographystyle{mn2e}

\end{document}